\documentclass[pre,aps,pacs,twocolumn]{revtex4}
\usepackage{epsfig}
\usepackage{bm}
\usepackage{color}
\newcommand{\B}[1]{{\bm{#1}}}

\usepackage[latin1]{inputenc}
\newcommand{\beq}{\begin{equation}}
\newcommand{\eeq}{\end{equation}}
\newcommand{\bea}{\begin{eqnarray}}
\newcommand{\eea}{\end{eqnarray}}

\begin{document}
\title{On the Anomalous Scaling Exponents in Nonlinear  Models of Turbulence}
\author{Luiza Angheluta$^{1,2}$, Roberto Benzi$^3$, Luca Biferale$^3$,  Itamar Procaccia$^1$ 
and Federico Toschi$^{4,5}$}
\affiliation{$^1$The Department of Chemical Physics, The Weizmann Institute of Science, Rehovot 76100, Israel. \\
$^2$ The Niels Bohr Institute, Blegdmasvej 17, Copenhagen.\\
$^3$ Dept. of Physics and  INFN, 
 University of Rome ``Roma Tor Vergata'', Via della Ricerca Scientifica 1, 00133 Rome, Italy.\\
 $^4$ Istituto per le Applicazioni del Calcolo CNR, Viale del Policlinico 137, 00161 Roma, Italy.\\
 $^5$ INFN, Sezione di Ferrara, Via G. Saragat 1, I-44100 Ferrara, Italy.}

\begin{abstract}
  We propose a new approach to the old-standing
  problem of the anomaly of the scaling exponents of nonlinear models
  of turbulence. We achieve this by constructing, for any given
  nonlinear model, a {\em linear} model of passive advection of an
  auxiliary field whose anomalous scaling exponents are {\bf the same}
  as the scaling exponents of the nonlinear problem.  The statistics
  of the auxiliary linear model are dominated by `Statistically
  Preserved Structures' which are associated with exact conservation
  laws.  The latter can be used for example to determine the value of
  the anomalous scaling exponent of the second order structure
  function. The approach is equally applicable to shell models and to
  the Navier-Stokes equations.
\end{abstract}
\maketitle
The calculation of the scaling exponents of structure functions of nonlinear
turbulent velocity fields remains one of the major open problems of
statistical physics \cite{Fri}. Dimensional considerations appear to fail to
provide the measured exponents, and present theory cannot even specify
the mechanism for the so called ``anomaly'', i.e. the deviation of the
scaling exponents from their dimensional estimates.  
Theoretical attempts to calculate the exponents were mainly based on
perturbative expansions \cite{00LP} or on closures of the infinite
correlation function hierarchy \cite{yakhot}.  The aim of
this Letter is to propose a new idea to ascertain the
anomaly of the scaling exponents in turbulence. In addition, we
exhibit an alternative way to determine the anomalous scaling exponent
of the second order structure function.  The proposed approach 
is equally applicable to Navier-Stokes turbulence and to simplified
models of turbulence, like nonlinear shell models.  The only
distinction is in the ease of numerical demonstration; for shell
models we present adequate numerical confirmation of the proposed
theory. For Navier-Stokes turbulence we present calculations at a
resolution of $128^3$.

The central idea is to construct a {\em linear} model whose scaling
exponents are {\em the same} as those of the {\em nonlinear}
problem. In this linear problem the exponents are universal to the
forcing, and we understand the mechanism for the anomaly of the
scaling exponents; we use this to show that also the {\em nonlinear}
problem must have anomalous exponents.  We exemplify the idea first in
the context of the Navier-Stokes equations. Consider a model for two
coupled vector fields $\B u$ and $\B w$,
\begin{eqnarray}
  \!\!\frac{\partial \B u}{\partial t}+\B u \cdot \B \nabla \B u+\lambda \B w \cdot 
  \B \nabla \B u &=& -\B \nabla p +\nu \nabla^2 \B u +\B f , \label{3}\\
  \!\!\frac{\partial \B w}{\partial t}+\B u \cdot \B \nabla \B w+\lambda \B w \cdot 
  \B \nabla \B w &=& -\B \nabla \tilde p \!+\nu \nabla^2 \B w +\B {\tilde f} , \label{4}  \nonumber
\end{eqnarray}
Here $\nu$ is the kinematic viscosity, $p$ and $\tilde p$ are pressure fields 
imposing $\B \nabla \cdot \B u  = \B \nabla \cdot \B w =0$ and
$\B f$ and $\B {\tilde f}$ are
two uncorrelated Gaussian random forcing. Finally $\lambda$ is a real number. Here and
below we {\em assume} that the scaling exponents are universal to the forcing.
We want to demonstrate their anomaly and to find their numerical values. 
For $\lambda=0$ Eq. (\ref{3}) reduces to the Navier-Stokes equations for $\B u$,
whereas Eq. (\ref{4}) becomes a linear equation for $\B w$, passively advected by $\B u$.
This linear problem was referred to before as a ``passive vector with pressure"  \cite{01AP,anto2,Benzi1}. It
is known to exhibit anomalous scaling with exponents that are  universal to the
forcing. In addition, one understand the  mechanism for the anomaly
\cite{CV,01ABCPM,02CGP,fgv}; the linear model possesses ``Statistically
Preserved Structures" (SPS) which are evident in the decaying problem.
These are {\em left} eigenfunctions of
eigenvalue 1 of the linear propagator for each order (decaying)
correlation function, and see below for more detail. 

Evidently, for any finite value of $0<\lambda<\infty$ the scaling
exponents of the two fields $\B u$ and $\B w$ must be the same, due to
the symmetry $ \lambda \B w \leftrightarrow \B u$ and the assumed
universality to the forcing.  Consider the two composite fields $\B
u_+ \equiv \B u+\lambda \B w$ and $\B u_- \equiv \B u - \lambda \B
w$. Choose the forcing terms in (\ref{3},\ref{4}) such that $\langle
(\B f+\lambda \B {\tilde f})(\B f-\lambda \B {\tilde f}) \rangle =
0$. Then $\B u_+$ satisfies precisely the Navier Stokes equations
(with forcing $\B f + \lambda \B {\tilde f}$) while $\B u_-$ is a
passive vector advected by $\B u_+$ and forced by $\B f -\lambda \B
{\tilde f}$.  Our main proposition is that the scaling exponents of
the Navier Stokes field $\B u_+$ and of the passive vector $\B u_-$
are the same. If true, we can study the anomalous scaling of the
Navier-Stokes problem by using the successful tools and the concepts
employed to understand the anomalous scaling for passive fields
(scalar or vector). Note that the identity of the scaling exponents of
$\B u_+$ and $\B u_-$ is equivalent to saying that in Eqs. (\ref{3})
and (\ref{4}) the scaling exponents of $\B u$ and $\B w$ are the same
for all $\lambda$ {\em including $\lambda=0$}.

\begin{figure}
\centering
\epsfig{width=.40\textwidth,file=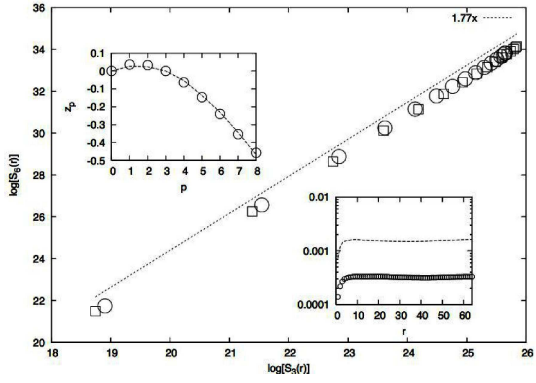}
\caption{ Log-log plot of the sixth 
order structure functions of the fields $\B u_+$ and $\B u_-$ (circles and squares respectively), for
$\lambda=1$, as a function of the 
third order structure functions. The dashed line corresponds to the best
fit in the scaling region with slopes $1.77$. Lower insert:  $S_6(r)/S_3(r)^{1.77}$. Upper insert:  $z_p \equiv \zeta_p/\zeta_3 -p/3 $ computed
for the structures functions of  $\B u_+$  (line) and  $\B u_-$ (circles).}
\label{figurens}
\end{figure}

To verify our proposition we present in Fig. (\ref{figurens}) 
the scaling properties of $\B u_+$ and $\B u_-$ for $\lambda=1$
obtained by direct numerical simulations of Eqs. (\ref{3}) and (\ref{4}).
The pseudospectral code used for the simulations has a resolution of $128^3$.  In Fig.  
(\ref{figurens}) we show the ESS plot of the sixth order structure functions for 
both $\B u_-$ and $\B u_+$. One observes convincing scaling behavior with the 
same exponent for both field. In the upper insert we show  the anomalous 
exponents $z_p \equiv \zeta_p/\zeta_3 -p/3$ for the field $\B u_+$ (line) 
and $\B u_-$ (circles) computed up to order $8$: the agreement is excellent. 
Thus, we can indeed propose  that the Navier-Stokes field  $\B u_+$ has the same scaling exponents as the 
passive vector field $\B u_-$. For additional strong evidence we turn to shell models \cite{GioBook,bif03,Gledzer}.

To reach a deeper understanding of the relation between
the nonlinear and the linear models, and to clearly present the role of the statistically Preserved
Structures, we consider the Sabra shell model which, like other shell models of turbulence,
is a truncated description of the dynamics of Fourier modes,
preserving some of the structure and conservation laws of the
Navier-Stokes equations:
\begin{eqnarray}
(\frac{d }{dt}&+&\nu k_n^2 ) u_n = i(k_{n+1} u_{n+1}^* u_{n+2}-\delta
k_n u_{n-1}^* u_{n+1}\nonumber\\&+&(1-\delta) k_{n-1} u_{n-1} u_{n-2})+f_n \ . \label{sabra}
\end{eqnarray}
Here $u_n$ are the velocity modes restricted to `wavevectors' $k_n=k_0
\mu^n$ with $k_0$ determined by the inverse outer scale of
turbulence. The model contains one additional parameter, $\delta$, and
it conserves two quadratic invariants (when the force and the
dissipation term are absent) for all values of $\delta$. The first is
the total energy $\sum_n |u_n|^2$ and the second is $\sum_n (-1)^n
k_n^{\alpha} |u_n|^2$, where $\alpha= log_{\mu} (1-\delta)$.  In this Letter we consider values of the parameters such
that $0<\delta<1$; in this region of parameters the second invariant
contributes only with sub-leading exponents to the structure functions
\cite{ditlev,sabra}. 
The scaling exponents characterize the
structure functions:
\begin{eqnarray}
S_2(k_n)\equiv \langle u_n u^*_n\rangle \sim k_n^{-\zeta_2} \ , \label{S2}\\
S_3(k_n)\equiv \Im \langle u_{n-1}u_n u_{n+1}^*\rangle \sim k_n^{-\zeta_3}\ ,\label{S3}\\
\text{etc. for higher order}~S_p(k_n)\sim k_n^{-\zeta_p} \  . \nonumber
\end{eqnarray}
The values of the scaling exponents were determined accurately by
direct numerical simulations. Besides $\zeta_3$ which is exactly
unity \cite{Piss93PFA}, all the other exponents $\zeta_p$ are anomalous, differing
from $p/3$. It was established numerically that the scaling exponents are universal, i.e. they are
independent of the forcing $f_n$ as long as the latter is
restricted to small $n$ \cite{bif03}.
Despite of the much simpler structure of Shell Models in comparison
with Navier-Stokes equations no systematic breakthroughs on the
analytical calculation of scaling properties has been
achieved. Previous attempts being manly based on stochastic closures
 \cite{parisi,Benzi2}.

Consider then a passive advected field which in the discrete shell
space has the complex amplitudes $w_n$. The dynamical equations for
this field are linear and constructed under the
following requirements: (i) the structure of the
  equations is obtained by linearizing the nonlinear problem and
  retaining only such terms that conserve the energy; (ii) the resulting equation is
identical with the sabra model when $w_n = u_n$; (iii) the energy is
the only quadratic invariant for the passive field in the absence of
forcing and dissipation. These requirements lead to the following
linear model:
\begin{equation}
\frac{dw_n}{dt} = \frac{i}{3}\Phi_n(u,w)-\nu k_n^2 w_n +f_n\ , \label{lin}
\end{equation}
where the advection term is defined as
\begin{eqnarray}
&&\Phi_n(u,w) = k_{n+1}[(1+\delta) u_{n+2}w_{n+1}^*+(2-\delta)u_{n+1}^*w_{n+2}]\nonumber\\
&&+ k_{n}[(1-2\delta)u_{n-1}^*w_{n+1}-(1+\delta)u_{n+1}w_{n-1}^*]\nonumber\\
&&+ k_{n-1}[(2-\delta)u_{n-1}w_{n-2}+(1-2\delta)u_{n-2}w_{n-1}]
\end{eqnarray}
Observe that when $w_n = u_n$ this model reproduces the Sabra
model, and also that the total energy is conserved because $\sum_n
\Im[\Phi_n(u,w)w^*_n] = 0$. The second quadratic invariant
is not conserved by the linear model. Finally, both models have the same `phase symmetry' in the
sense that the phase transformations $u_n\to u_n \exp{(i\phi_n)}$ and $w_n\to w_n\exp{(i\theta_n)}$ 
leave the equations invariant iff
$\phi_{n-1}+\phi_n = \phi_{n+1}$, $\theta_{n-1}+\theta_n=\theta_{n+1}$.
This identical phase relationship
 guarantees that the non-vanishing correlation functions of both
models have precisely the same forms. 
Thus for example the only  second and third
correlation functions in both models are those written explicitly in Eqs. (\ref{S2}) and (\ref{S3}).

As already remarked, the anomalous scaling of $w_n$ can be investigated in terms
of the SPS \cite{fgv,01ABCPM,02CGP}. For example for the second order
correlation function denote the propagator $P^{(2)}_{n,n'}(t|t_0)$;
this operator propagates any initial condition $\langle w_n
w^*_n\rangle (t_0)$ (with average over initial conditions, independent
of the realizations of the advecting field $u_n$) to the decaying
correlation function (with average over realizations of the advecting
field $u_n$)
\begin{equation}                                  
\langle w_n w^*_n\rangle(t) = P^{(2)}_{n,n'}(t|t_0)\langle w_{n' }w^*_{n'}\rangle (t_0)  \ .
\end{equation}
The second order SPS, $Z^{(2)}_n$, is the left eigenfunction with eigenvalue 1,
\begin{equation}
Z^{(2)}_{n'} = Z^{(2)}_n P^{(2)}_{n,n'}(t|t_0) \ .
\end{equation}
Note that $Z^{(2)}_n$ is time independent even though the operator
$P^{(2)}_{n,n'}(t|t_0)$ is time dependent. Each order correlation
function is associated with another propagator $\B P^{(p)}(t|t_0)$ and
each of those has an SPS, i.e. a {\em left} eigenfunction $\B Z^{(p)}$
of eigenvalue 1. These non-decaying eigenfunctions scale with $k_n$,
$\B Z^{(p)}\sim k_n^{-\xi_p}$, and the values of the exponents $\xi_p$
are anomalous. Finally, 
one can show that  these SPS  are also the leading scaling
contributions to the structure functions of the {\em forced} problem (\ref{lin})  \cite{fgv,01ABCPM}.
Thus {\bf the scaling exponents of the linear problem are independent
of the forcing $f_n$}, since they are determined by the SPS of the
decaying problem.
\begin{figure}
\centering
\epsfig{width=.40\textwidth,file=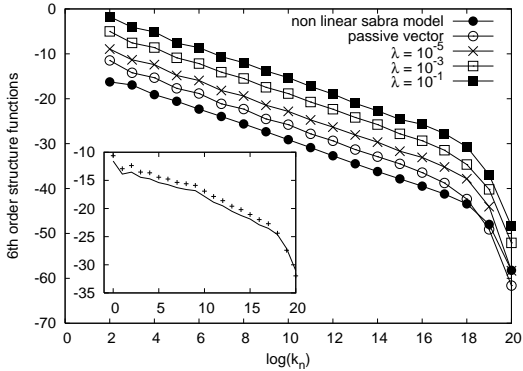}
\caption{ The sixth 
order structure function of the field $w_n$ in Eqs. (\ref{2}) for
$\lambda=10^{-1}, 10^{-3}$ and $10^{-5}$, 
together with the sixth order structure function for the
Sabra model (\ref{sabra})  and 
for the linear model (\ref{lin}), respectively. 
The structure function of the field $u_n$ for $\lambda>0$ are not shown since they are indistinguishable from those of the $w_n$.
Inset:  log-log plot of the fourth-order correlation function
$F_{2,2}(k_n,k_7)$ vs. $k_n$ calculated for the linear field ($+$) and for
the nonlinear field (solid line)  at $\lambda=0$.}
\label{lambdato0}
\end{figure}

Using equations (\ref{lin}), we can now write the (Sabra) shell model version of Eqs. (\ref{3}),(\ref{4}):
\begin{eqnarray}
\frac{du_n}{dt} &=& \frac{i}{3}\Phi_n(u,u)+\frac{i \lambda}{3}\Phi_n(w,u)-\nu k_n^2 u_n +f_n\ , \label{1}\\
\frac{dw_n}{dt} &=& \frac{i}{3}\Phi_n(u,w)+\frac{i \lambda}{3}\Phi_n(w,w)-\nu k_n^2 w_n +\tilde f_n \label{2}
\end{eqnarray}
For $\lambda=0$ we recover
the equations for the nonlinear and a linear models, 
Eqs. (\ref{sabra}) and (\ref{lin}). At this point we present strong evidence
that the scaling exponents of either
field exhibits no jump in
the limit $\lambda\to 0$.  Accordingly,  the  scaling exponents of
either field can be  obtained from the SPS of the linear
problem.

Eqs. (\ref{1}) and (\ref{2}) were solved numerically, choosing $f_n$ a
constant complex number limited to $n=0,1$, and $\tilde f_n$ a random
force with zero mean, operating on the same shells.  We chose
$\nu=10^{-8}$, $\delta = 0.6$ and $\lambda=10^{-1},
10^{-3},10^{-4},10^{-5},0$. 
In Fig. \ref{lambdato0} we show, for
example, results for the sixth order objects $\langle |u_{n-1}u_n
u_{n+1}^*|^2 \rangle$ and $\langle |w_{n-1}w_n w_{n+1}^*|^2 \rangle$.
Plotted are double-logarithmic plots of these object as a function of
$k_n$. We see that the exponents of the linear and nonlinear model at 
 $\lambda =0$  are the same and they do coincide with the exponents
 of the two coupled models (\ref{1}), (\ref{2}) for   $\lambda>0$. The same
results have been checked numerically up to exponents of order $10$.
Hence, the limit $\lambda\to 0$ is regular. 
  
We stress at this point that the two problems do not
share {\em exactly} the same statistics; the linear problem, being
symmetric in $w_n\to -w_n$ has an even probability distribution
function (pdf) and thus zero prefactors for all the odd structure
functions. The statement is only
about the identity of the scaling exponents, neither the trajectory
in phase space nor the the pdf. 
%
%
In the inset of Fig. \ref{lambdato0}
we also demonstrate that the linear and the nonlinear problems share
the same scaling properties for correlations that depend on more than one shell. 
The data pertain to $F_{p,q}(k_n,k_m) \equiv \langle |u_n|^p
|u_m|^q \rangle $, with $p=2,q=2$ for both models.
Finally, we comment
that the limit $\lambda\to 0$ can be considered mathematically for the
shell model equations (\ref{1}) and (\ref{2}), to prove that it is 
not singular. Such a proof is however beyond the scope of this
Letter, and will be presented elsewhere.
\begin{figure}
\centering
\epsfig{width=.45\textwidth,file=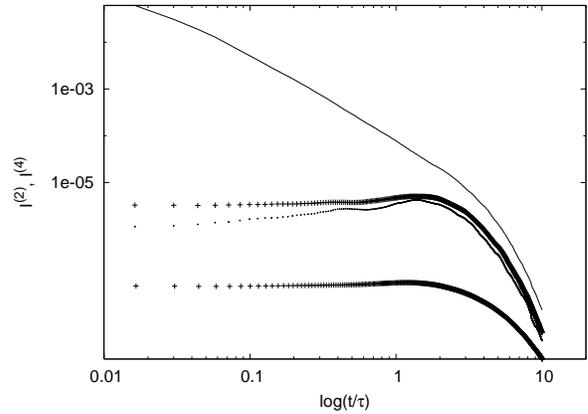}
\caption{With the symbols ($+$)
the constants $I^{(2)}$ (bottom) and $I^{(4)}$ (top) constructed by
  projecting the decaying structure function of the linear model on  the {\em forced} structure function of the {\bf nonlinear} model.  To emphasize the importance of using the correct SPS, we 
  also show the result for  $I^{(4)}$  using the dimensional Kolmogorov 
prediction for $Z^4$ (small dots) and $Z^4=1$ (solid line)  }
\label{I2}
\end{figure}

The greatest asset of the present approach is that we can now 
forge a connection between the SPS of the linear model and  
the {\em forced} correlation function of the nonlinear problem. This 
underlines the anomaly of the scaling properties of the latter model, and allows us to determine
$\zeta_2$. We start with the second order quantities. 
We can project a generic second order {\bf decaying} correlation function
of the linear model onto the second order SPS, thus creating a
statistically conserved quantity:
\begin{eqnarray}
I^{(2)} &\equiv& \sum_n Z^{(2)}_n \langle w_n w^*_n\rangle(t) = 
\sum_{n,n'}Z^{(2)}_nP^{(2)}_{n,n'}(t|t_0)\langle w_{n' }w^*_{n'}\rangle (t_0)\nonumber\\
&=&\sum_{n'}Z^{(2)}_{n'}\langle w_{n' }w^*_{n'}\rangle (t_0) \ . \label{const}
\end{eqnarray}
Where the average is  over different initial conditions for
the linear fields  and
different realization  of the advecting velocity field. To show that the forced second order
correlation of the nonlinear field is dominated by $Z^{(2)}$, we use this forced 
correlation function {\em instead of} $Z^{(2)}$ in Eq. (\ref{const}). The test is whether 
 $I^{(2)}$ remains constant on a time window which increases with Reynolds. This is shown in
Fig.~\ref{I2}.  The success of this test  demonstrates that (i) there exists a SPS
for the linear problem; (ii) the SPS is well represented by the {\it
forced nonlinear} second order correlation functions. 
This is a direct demonstration that the
correlation function of the nonlinear model 
scales with the same anomalous exponent as $Z^{(2)}$. 
An even more stringent test can be made using SPS of orders
large than 2, where also correlations between different shells are
relevant for the decaying properties \cite{01ABCPM,02CGP}.  For
example $I^{(4)}$ is given by the weighted sum of three contributions:
\begin{eqnarray}
\label{eq:i4}
I^{(4)} =& \sum_{n,m}& Z_{n,m}^{(a,4)} \langle |w_n|^2 |w_m|^2 \rangle(t) + \\
 &\sum_{n}&[ Z_{n}^{(b,4)} \langle w_nw^2_{n+1}w^*_{n+3}  \rangle(t)  +c.c. ] + \nonumber \\
&\sum_{n}& [Z_{n}^{(c,4)} \langle w_nw_{n+1}w_{n+3}w^*_{n+4}  \rangle(t) +c.c.] \nonumber\ ,
\end{eqnarray}
where all the terms allowed by the phase
symmetry  were employed.  In Fig.~\ref{I2} we show results
for $I^{(4)}$ where again we swapped the SPS of the linear problem for the measured {\em forced} correlations of the nonlinear problem: 
$Z_{n,m}^{(a,4)} \rightarrow \langle |u_n|^2 |u_m|^2 \rangle$ and the
corresponding expressions 
for $Z_n^{(b,4)}$ and $Z_n^{(c,4)}$.
We thus conclude that the scaling exponents of a given nonlinear shell
model can be understood from the SPS of
an appropriately constructed linear problem. To make this point
crystal clear, we have used in fact the forced structure functions of
the nonlinear model as  approximants for $Z^{(2)},Z^{(4)}$ in the calculation
of $I^{(2)}$ and $I^{(4)}$ shown in Fig. \ref{I2}. The constancy of
both demonstrates that the forced correlation function of the
nonlinear model are very well approximated by the SPS of the linear
model. This demonstration can be repeated with higher order
correlation functions with the same (or better) degree of
success. Finally,  the existence of a conserved quantity $I^{(2)}$
can be used to {\it calculate} $\xi_2=\zeta_2$. 
Starting from a given arbitrary initial condition (say a $\delta$-function on one shell) and
computing  Eq. (\ref{const}) with many realizations of the
advecting velocity field, one finds that there exists a {\em sharply defined} $\xi_2$, $ Z^{(2)}_n\sim
k_n^{-\xi_2}$, for which $I^{(2)}$ is indeed constant. The same approach 
can be used to determine $\zeta_3$ but we know that $\zeta_3=1$. Unfortunately, this simple
approach cannot be used for higher order exponents, because the
corresponding SPS depend on more than one $k_n$, and cannot be represented as a simple
power law.

In conclusion the
anomalous scaling of nonlinear modes of turbulence, be then the Navier-Stokes equations
or shell models, are fixed by the
eigen functions of the inertial operator, which are precisely the SPS of the linear
problem. Thus, although the concept of eigenfunctions cannot be applied directly
in nonlinear problems, we are able to argue that the mechanism leading
to anomalous scaling in Navier-Stokes equations and other nonlinear
models is identical to the one recently discovered for the case of
passively advected fields.  This conclusion may open the way to a
deeper understand of intermittency in turbulent flows and to a direct
computation of the anomalous exponents.

We acknowledge useful discussion with J.-P. Eckmann, U. Frisch and M. Vergassola. LA is grateful to M.H. Jensen for encouragements and useful discussions. This work has been supported in part 
by the European Commission under a TMR grant.

\end{document}